# Proposed experiment of which-way detection by longitudinal momentum transfer in Young's double slit experiment

Masanori Sato
*Honda Electronics Co., Ltd.,*
*20 Oyamazuka, Oiwa-cho, Toyohashi, Aichi 441-3193, Japan*
E-mail: msato@honda-el.co.jp

**Abstract**  The momentum of a photon may reveal the answer to the "which way" problem of Young's double slit experiments. A photon passing through the boundary between two media, in which a photon travels at different velocities, undergoes a momentum change according to the law of conservation of momentum. The momentum of the photon is transferred locally to the medium, and the boundary between the media receives stress, which determines the photon trajectory. An experiment is performed using a crystal plate that can transform the stress to electric charge. We are able to detect the electric charge after the detection of the photon on the screen, and control the sensitivity of photon detection. By means of this proposed experiment it is determined whether or not an attempt to detect the "which way" of photon travel destroys the interference patterns.

**1.  Introduction**

Heisenberg's uncertainty relations and complementarity have been discussed previously [1-12] as being the reasons why it has not been possible to determine the direction in which a photon travels in Young's double slit experiments [1-3]. The destruction of interference patterns can be explained in terms of uncontrolled momentum kicks to the particle [13]. In recent years, Scully and co-workers [4, 5] have claimed that their scheme destroys the interference patterns without transferring any transverse momentum to the particle. It is said that an attempt to detect the direction of photon travel will destroy the interference patterns, because of complementarity and uncertainty. Two groups have been debating the relative priority of complementarity [4, 5] and uncertainty [6-9]. The local momentum transfer (which occurs the particle is localized at a single slit) versus non-local momentum transfer (which if the particle is delocalized at both slits) has been discussed [13]. It is also discussed that the Wignerian analysis and the Bohmian analysis clearly distinguish between a local momentum transfer and a nonlocal momentum transfer [14].

Active measurement, for example, the electromagnetic stimulation of electrons [4,5], destroys interference patterns, but we think that passive measurement, for example, local momentum transfer, does not destroy the interference patterns. Therefore, there is a possibility of detecting simultaneously the trajectory and interference patterns, if we detect the photon trajectory by the longitudinal momentum.

In this letter, the momentum discussed is longitudinal rather than transverse, which simplifies the argument. Therefore, a newly proposed Young's double slit interference experiment is presented and the method of detection of "which way" of photon travel in the interference experiment is described. In this experiment, classical and local momentum kicks to the particle can be controlled linearly. Therefore, we are able to test whether or not the local momentum transfer destroys the interference patterns.

## 2. Momentum conservation law and stress

In this experiment, the photon momentum $\hbar k$ is used. Here, k is the wave number and $\hbar = \dfrac{h}{2\pi}$ (h is Planck's constant). A photon changes its momentum according to the medium though which it is propagating. It also changes at the boundary between the media, resulting in a pressure increase there. A photon has energy $\hbar \omega$ and momentum $\hbar k$; here ω is the frequency. The velocity of a photon changes according to the medium through which the photon travels, as do the wave number and the momentum. The transfer of the momentum difference to the medium is the cause of the pressure increase at the boundary.

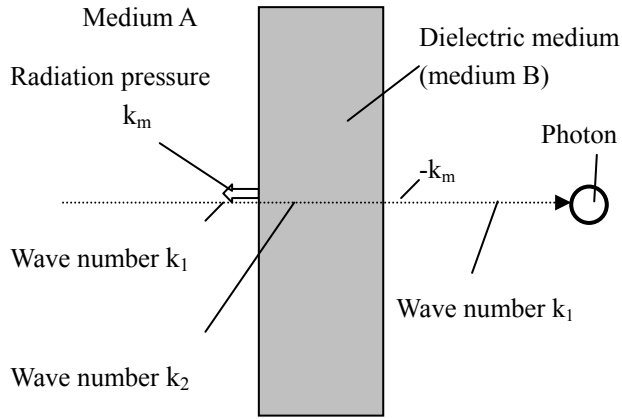

**Fig. 1** Radiation pressure depends on the speed of light
The difference in the wave number (the speed of light) causes radiation pressure at the surfaces of the two media.

**Figure1** shows the local momentum transfer of a photon that propagates from medium A to medium B, and the stress at the boundary. The law of conservation of momentum is

$$k_1 = k_2 + k_m. \qquad (1)$$

Here, $k_1$ is the photon wave number in medium A, $k_2$ is the wave number in medium B, and $k_m$ is the wave number corresponding to the momentum transferred to the medium. The momentum transfer of equation (1) is a local phenomenon. The directions of $k_1$, $k_2$ and $k_m$ are normal to the surface, and the momentum transferred to the medium generates pressure (or stress) at the boundary between the two media. The trajectory of the photon is manifested in the medium as pressure.

## 3. Proposed Young's double slit experiment

In this proposed experiment, the detection of the momentum transferred locally to the medium is discussed. **Figure 2** shows an experimental setup, using AT cut crystal plates with transparent electrodes. If a photon passes through the crystal plate, the stress on the surface induces charges on the electrodes. We can measure the charges by means of a switch and a current meter, after photon detection on the screen. It is already known that the interference pattern appears if the charges are not detected. Of course, it is impossible to determine whether or not a photon passes through the crystal plate, because the signal is hidden in thermal noise. However, we can determine whether or not the attempt of detection destroys the interference patterns. In this experiment, a photon loses some energy due to the detection of the charges, so the redshift $\omega_{red}$ occurs. The law of conservation of energy is

$$\omega_1 = \omega_2 + \omega_{red} + \Delta\omega_{red} \quad (2)$$

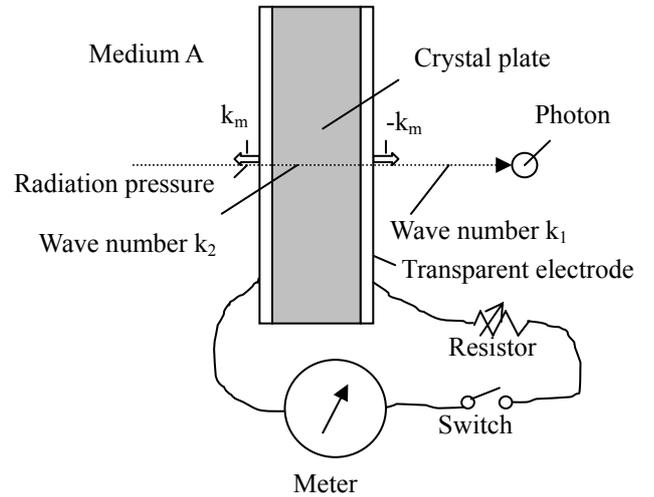

**Fig. 2** Frequency redshifter using a AT-cut crystal plate
The stress on the surface induces the charges. We could detect the charges after the photon arrived at the screen by using the switch. The sensitivity of photon detection is controlled using the variable resistor.

Here, $\omega_1$ is the photon frequency in medium A, $\omega_2$ is the frequency in medium B, $\omega_{red}$ is the redshift (i.e., discharged energy), and $\Delta\omega_{red}$ is the deviation of the redshift. We assume that the detection of the charges causes an uncontrolled momentum kick to a photon, which is presented as the deviation of the redshift $\Delta\omega_{red}$. Here, the detection of the charges is performed by the discharge that is disturbed by thermal noise, which causes deviation of the redshift $\Delta\omega_{red}$. However, we can modify the redshift by controlling the sensitivity of the charge detection. The sensitivity is controlled using a variable resistor in the circuit. At a high resistance the charges are not completely discharged and therefore we cannot obtain the required information, but at lower resistance, we can obtain complete information.

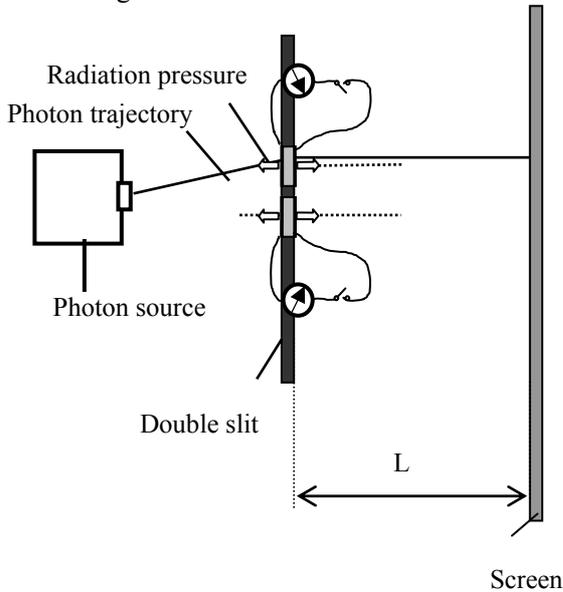

**Fig. 3** Proposed Young's double slit experiment

The setup for the experiment is shown in **Fig. 3**. If we place the crystal plates as dielectric media in the path of a photon, one of the dielectric media will obtain momentum according to the passage of the photon. In this argument, two paths are observed with distinct separation. According to the law of conservation of momentum, the trajectory of a photon, which is indicated by the pressure on the boundary of the dielectric media, is shown. Therefore, the proposed experiments test whether or not the detection of photon energy, which is equivalent to the stress, destroys the interference patterns.

With this experiment, we can test the correlation between the path detection and the interference pattern generation. The path detection procedure may result in a redshift, which is controlled by the sensitivity of the circuit. Therefore, we have two methods with which to control the interference pattern, one by controlling the sensitivity of the circuit, and the other by controlling the distance between the crystal plate and the screen, which is indicated by L in **Fig. 3**, because the redshift $\omega_{red}$ and the distance L cause the phase shift of a photon on the screen.

## 4. Application of the Heisenberg relations

We can testify this argument, using the Heisenberg relations for the position and the momentum of a photon, which is represented

$$\Delta x \Delta p \geq \frac{h}{4\pi}, \quad (3)$$

where x is the position, p is the photon momentum, and $\Delta$ means a deviation. We discuss the condition that a photon has the position and the momentum, using the Heisenberg relations.

Here we discuss the momentum $p_m$, which is the momentum that is transferred to the medium, and if the light speed is 20% decreased in the dielectric medium

$$p_m \approx \frac{h}{\lambda} \times 0.2 \quad ( p_m \approx \frac{p}{5} ), \quad (4)$$

here $\lambda$ is wavelength. If we try to obtain the momentum $p_m$ with a 10% deviation. Then assuming

$$\Delta p_m \approx 0.1 \times p_m = \frac{p}{50}, \qquad (5)$$

we obtain the condition of the deviation for position

$$\Delta x \geq \frac{50\lambda}{4\pi}. \qquad (6)$$

Equation (5) shows that we need the distance, at least, equal to the deviation $\Delta x = \frac{50\lambda}{4\pi}$, when we try to measure the momentum $p_m$ with a 10% deviation. So if we take

$$x_1 \geq \Delta x \geq \frac{50\lambda}{4\pi}, \qquad (7)$$

where, $x_1$ is the distance between the two slits. This condition will be fulfilled in this experiment, which guarantees that we can distinguish the momentum $p_m$ theoretically. It indicates that we could distinguish the "which way" by the momentum $p_m$, if we take the distance $x_1$ in Eq.(8). For example, using the 630nm laser, the distance between the two slits $x_1 \geq 2.5$ μm is enough to distinguish which way a photon has passed.

In this proposed experiments, these discussions indicate that the mechanism which interferes to detect the "which way" of photon travel is not the Heisenberg relations but thermal noise disturbance.

**5. Discussions**

In this letter, we argued the detection of the "which way" of photon travel, but we did not discuss the mechanism of the interference pattern generations. We think Bohm's quantum potential [15], which simultaneously determine the photon trajectories and interference patterns, is one of the most suitable theoretical backgrounds of this experiment.

We think the feasibility of these experiments depends on the technology of thermal noise reduction. We do not think the feasibility of these experiments are not restricted by the Heisenberg relations.

**6. Conclusions**

In this letter, we proposed a feasible experiment, which demonstrates the possibility of simultaneously determining the photon trajectory and the interference pattern. In this experiment, if the switches are off and the measurement is not performed, the interference patterns can be detected. However, the sensitivity of photon detection can be controlled using a variable resistor. Of course, photon detection by the crystal plate cannot be performed, because of the thermal noise disturbance, but not the Heisenberg relations. However, we can determine whether the attempt to detect the photon trajectory destroys the interference patterns, and whether the phase shift of the photon according to the frequency redshift $\omega_{red}$ and the deviation $\Delta\omega_{red}$ can be restricted by adjusting the distance L. Parameters such as the sensitivity of photon detection and the distance L can be modified linearly. In this experiment, longitudinal rather than transverse momentum is discussed. In other words, the correlation between the detection of the photon trajectory and interference pattern generation is discussed, i.e., we do not discuss the relation between transverse momentum transfer and interference pattern generation. We could obtain information on complementarity and uncertainty with regard to the "which way" problem of Young's double slit experiments.